\documentclass[preprint2]{aastex}

\usepackage{xcolor}
\usepackage{subfigure}
\usepackage{amsmath}

\slugcomment{accepted by ApJ, July 18th 2016}

\shorttitle{Vortices and spirals in HD135344B}
\shortauthors{van der Marel et al.}

\begin{document}


\title{Vortices and spirals in the HD135344B transition disk}



\author{N. van der Marel}
\affil{Institute for Astronomy, University of Hawaii, 2680 Woodlawn dr., 96822 Honolulu, HI, USA}
\email{marel@hawaii.edu}

\and
\author{P. Cazzoletti} 
\affil{Max-Planck-Institut fur Extraterrestrische Physik, Giessenbachstrasse 1, 85748 Garching, Germany}

\and
\author{P. Pinilla}
\affil{Leiden Observatory, P.O. Box 9513, 2300 RA Leiden, the Netherlands}

\and
\author{A. Garufi}
\affil{ETH, Zurich, Wolfgang-Pauli-Strasse 27, 8093, Zurich, Switzerland}
\altaffiltext{1}{Universidad Auton\'{o}noma de Madrid, Dpto. F\'{i}sica Te\'{o}rica, M\'{o}dulo 15, Facultad de Ciencias, Campus de Cantoblanco, E-28049 Madrid, Spain}




\begin{abstract}
In recent years spiral structures have been seen in scattered light observations and signs of vortices in millimeter images of protoplanetary disks, both probably linked with the presence of planets. We present ALMA Band 7 (335 GHz or 0.89 mm) continuum observations of the transition disk HD135344B at unprecedented spatial resolution of 0.16", using superuniform weighting. The data show that the asymmetric millimeter dust ring seen in previous work actually consists of an inner ring and an outer asymmetric structure. The outer feature is cospatial with the end of one of the spiral arms seen in scattered light, but the feature itself is not consistent with a spiral arm due to its coradiance. 
We propose a new possible scenario to explain the observed structures at both wavelengths. Hydrodynamical simulations show that a massive planet can generate a primary vortex (which dissipates at longer timescales, becoming an axisymmetric ring) and trigger the formation of a second generation vortex further out. Within this scenario the two spiral arms observed at scattered light originate from a planet at $\sim$30 AU and from the secondary vortex at $\sim$75 AU rather than a planet further out as previously reported.

\end{abstract}



\keywords{instabilities --- protoplanetary disks --- planets and satellites: formation --- planet-disk interactions}



\section{Introduction}
Protoplanetary disks are the cradles of young planets, where several dynamical processes are likely involved in the planet formation process \citep[e.g.][]{Armitage2011}. Of particular interest are the transition disks, disks with inner millimeter-dust cavities. In the last decade, observations have revealed that {some} transition disks are far from axisymmetric: azimuthal asymmetries in the submillimeter continuum are thought to be dust traps, triggered by  vortices acting as azimuthal pressure bumps \citep[e.g.][]{vanderMarel2013,LyraLin2013,Birnstiel2013}. On the other hand, near-infrared scattered light observations show large spirals \citep[e.g.][]{Muto2012,Garufi2013,Grady2013,Avenhaus2014}. Both spirals and vortices {may} indicate the presence of recently formed massive planets: in the case of a vortex through Rossby wave instability (RWI) {at the steep edges} of the gap that is carved by the planet \citep[][]{Lovelace1999,deVal-Borro2007} and in the case of  spirals through the trigger of density waves directly by the planet {\citep[e.g.][]{Kley2012}}.

\begin{figure*}[!ht]
\begin{center}
\epsscale{2}
\plotone{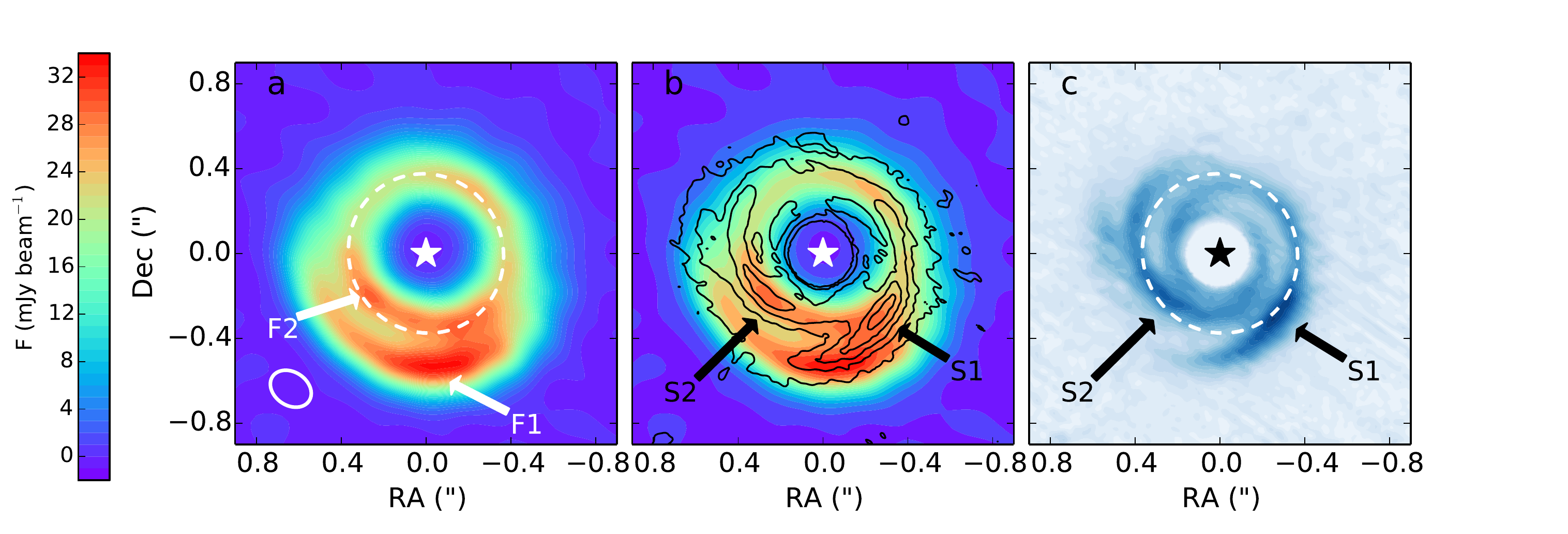} 
\end{center}
\caption{{\bf a.} 335 GHz continuum emission of HD~135344B in superuniform weighting. 
{\bf b.} Overlay of the PDI image of \citet{Garufi2013} (black contours) on top of the ALMA continuum emission. The spirals as identified by \citet{Muto2012} are labeled as S1 and S2. {\bf c.} PDI image of \citet{Garufi2013} in blue colors. In a and c, the white dashed ellipse indicates the 45 AU radius.
\label{fig:imagedata}}
\end{figure*}

Alternative explanations for spiral arms in disks include RWI at the edge of a dead zone \citep{lyra2015}, accretion from an envelope \citep{lesur2015} and gravitational instability \citep{Lodato2004,Lodato2005,Rice2004}, although  estimated disk masses generally appear to be too low for them to be self-gravitating \citep{WilliamsCieza2011}. 

A natural question is whether there is any relation between the spiral arms observed in near-infrared scattered light (from the disk surface) and the structures seen in submillimeter emission (from the midplane). Although spiral features in submillimeter emission have been seen in two transition disks \citep{Pietu2005,Christiaens2014}, they are not entirely consistent with their near infrared counterparts. \citet{Juhasz2015,Pohl2015,Dong2015spirals} demonstrated that spirals generated by planet-disk interaction more likely results from changes in the vertical structure rather than the density structure, which are hard to detect in millimeter emission. On the other hand, spirals that form through gravitational instability can trap dust \citep{Lodato2004,DiPierro2015,Dong2015GI}, resulting in millimeter continuum spirals. 

In this Letter we present Atacama Large Millimeter/submillimeter Array (ALMA) submillimeter continuum observations at very high spatial resolution of HD~135344B \footnote{also known as SAO~206462} (F4 star, $d\sim$140 pc, $\sim$8 Myr \citep{vanBoekel2005,Grady2009}), a well-studied transition disk at both optical and millimeter wavelengths. The HD~135344B disk contains a $\sim$40 AU radius dust cavity \citep{Brown2007,Brown2009,Andrews2011} with a minor azimuthal asymmetry along the dust ring \citep{Perez2014,Pinilla2015beta}. CO observations and scattered light indicate that gas and small grains are present inside the cavity \citep{Pontoppidan2008,Lyo2011,vanderMarel2015-12co,vanderMarel2015-isot,Garufi2013}, consistent with a scenario where a massive planet at $\lesssim$30 AU has cleared its orbit and trapped the large dust further out \citep{Pinilla2012b}. Scattered light imaging reveal two major spiral arms \citep{Muto2012, Garufi2013, Stolker2016}, proposed to be linked to planet-disk interaction, with planets located at  55 and 126 AU. 

The new images presented in this letter show substructure in the millimeter emission to an unprecedented level, revealing a double structure, which may be responsible for triggering the spiral arms seen in the scattered light. This new interpretation has consequences for the implied location of the putative planets.

\section{Observations}
HD~135344B was observed in ALMA Cycle 1 program 2012.1.00158.S (PI van Dishoeck) in Band 7 ($\sim$335 GHz or 896 {$\mu$m}) in the C32-5 configuration (20-800 m baselines), previously presented in \citet{vanderMarel2015-isot,Pinilla2015beta}. The spectral settings and calibration are discussed in \citet{vanderMarel2015-isot}. For this work, the continuum emission is re-imaged using superuniform rather than briggs weighting of the observed visibilities, resulting in a smaller beam size of 0.20$\times$0.16'' (Figure \ref{fig:imagedata}a). In superuniform weighting, the weights of the grid cells in the $u,v$-plane are set inversely proportional to the sampling density function, minimizing the sidelobes over an arbitrary field size,  whereas briggs weighting sets the weights also inversely proportional to the noise variance of each visibility. The peak S/N decreases from 210$\sigma$ (briggs) to 120$\sigma$ (superuniform) with $\sigma$ the rms level (0.25 mJy beam$^{-1}$). We also make use of archival data of HD135344B obtained in Polarization Differential Imaging (PDI) in the K$_s$ band ($\sim$2.2 micron) \citep{Garufi2013} with VLT/NACO. The data thus obtained trace the (polarized) scattered light from the disk surface and have angular resolution of 0.09''.

Figure \ref{fig:imagedata}a reveals that the millimeter emission does not originate from a single dust ring with an azimuthal asymmetry, but an outer azimuthal asymmetric feature in the south (labeled F1) and an inner ring-like feature F2. With the current spatial resolution it remains unclear whether they are connected in the south-west. These features are located at 45 and 75 AU radius. The F1 feature is at least 4 times brighter than its opposite side in the north, while the F2 ring is almost azimuthally  symmetric, with an azimuthal contrast of at most a factor 1.2. The peak brightness temperature is 20 K, implying that the emission is optically thick even at this wavelength (896 $\mu$m).

Figure \ref{fig:imagedata}b shows the overlay of the PDI image (multiplied by the squared distance to the central star) on top of the ALMA data. The ALMA features appear to follow the spiral structure: F1 is at the end of the spiral S1 (as defined in \citet{Muto2012}) while F2 appears to  overlap with S2. The brightest part of the S1 spiral in the west is however not cospatial with the brightest ALMA data points, and as we will show below S1 and F1 are related in a different way. 

In the modeling, we use the stellar position 15$^h$15$^m$48.$^s$42 -37$^{\circ}$09'16.''36) as derived from the $^{13}$CO emission, and for the deprojection a position angle of $62^{\circ}$ and an inclination of $16^{\circ}$ \citep{vanderMarel2015-isot}.



\section{Morphology}

\begin{figure*}[!ht]
\epsscale{1}
\plotone{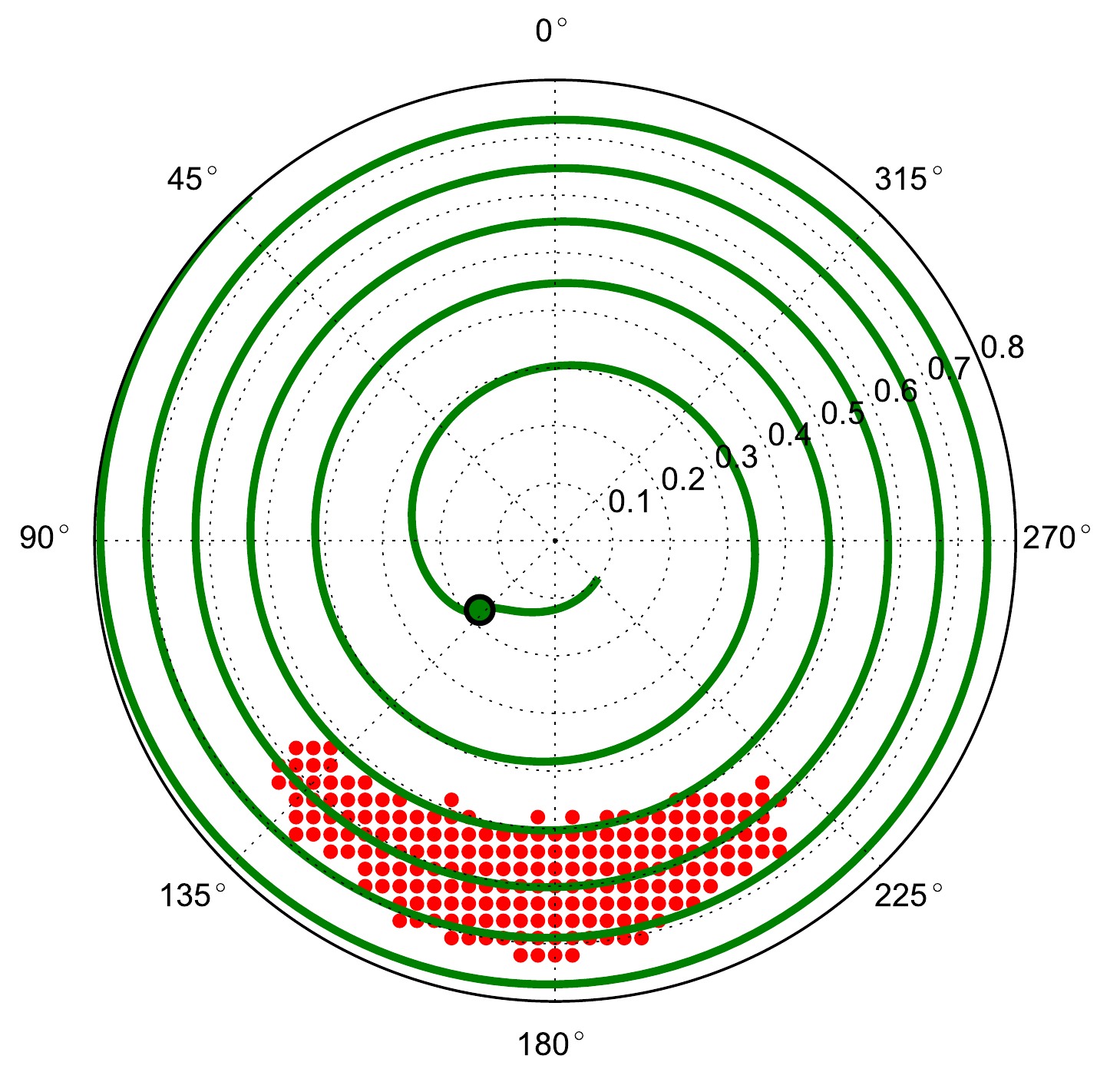}
\plotone{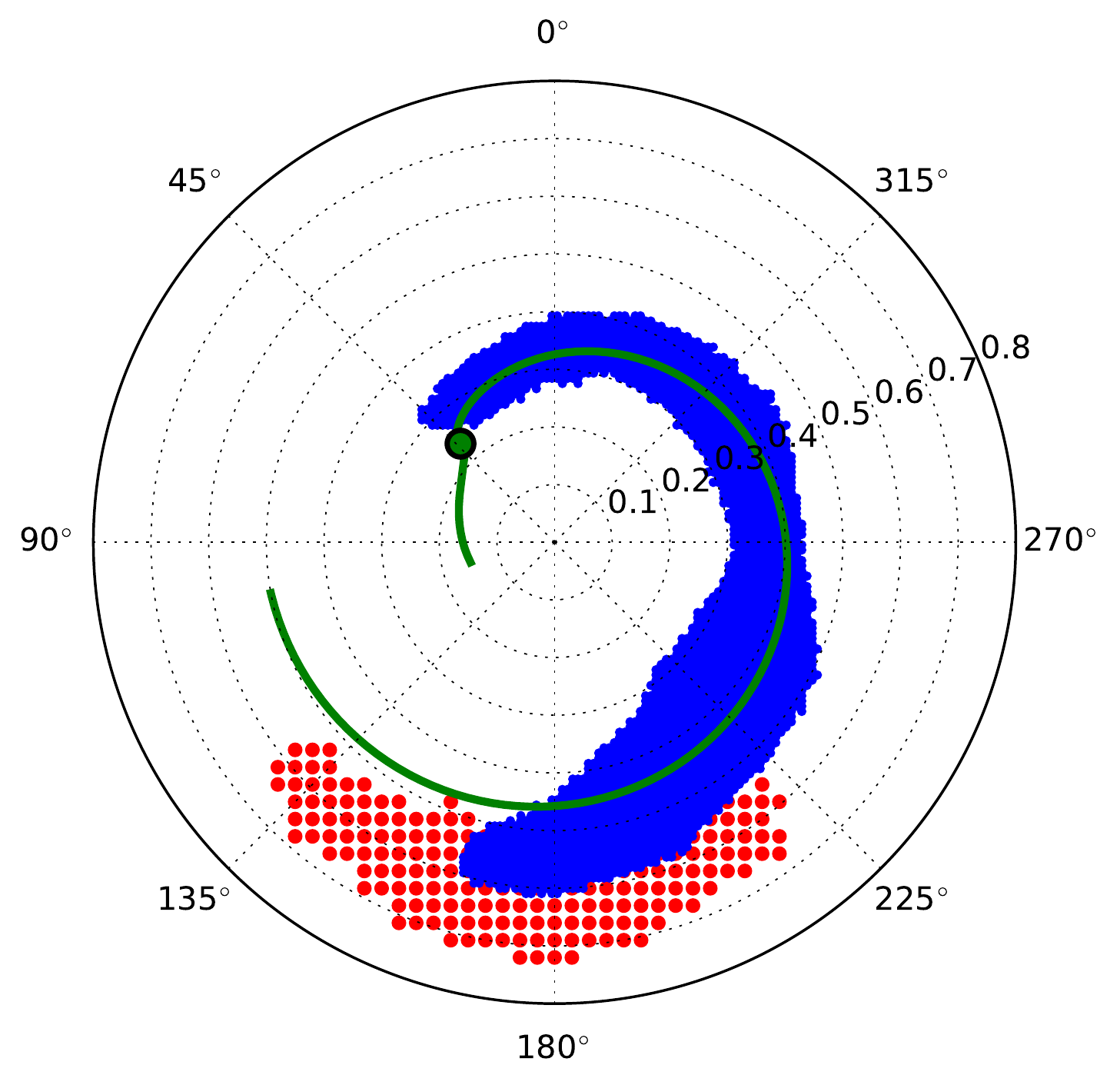}
\caption{{\bf Left.} Best fit of the ALMA continuum F1 feature (red dots) to the spiral model (green). This fit is unrealistic, as density waves are damped efficiently and the fourth winding would no longer be visible.
{\bf Right.} Overlay of the brightest data points of ALMA continuum (red dots) and the S1 feature in the scattered light data (blue dots, \citet{Garufi2013}), both deprojected. The green line shows our best-fit spiral to the blue data points, with $(r_c,\theta_0)=(0.24",134^{\circ})$. This figure shows that F2 does not follow the spiral arm seen in scattered light.}
\label{fig:spiraloverlay}
\end{figure*}

In order to understand the morphology of the disk, two different models are investigated. Model 1 follows the spiral description derived by \citet{Muto2012}. Model 2 consists of an inner symmetric ring and outer azimuthal asymmetry, following the morphology of the image. This double structure has been seen in certain 2D hydrodynamical simulations of planet disk interaction, with a primary vortex at the outer edge of the planetary gap and subsequently, a second vortex external to the primary \citep{Lobo-Gomes2015}. 

\subsection{The Spiral Model}
We model the shape of a spiral density wave generated by a planet located at $(r_{\rm c},\theta_0)$ using the analytical shape derived by \cite{rafikov2002}. This analytical approach describes the propagation of the wave from a launching point and it is given by 

\begin{equation}\label{eq:spiral}
\begin{split}
\theta(r)&=\theta_0-\frac{\rm sgn(r-r_{\rm c})}{h_{\rm c}}\\
&\times \Bigg[\bigg(\frac{r}{r_{\rm c}}\bigg)^{1+\beta}\Bigg\lbrace \frac{1}{1+\beta}-\frac{1}{1-\alpha+\beta}\bigg(\frac{r}{r_{\rm c}}\bigg)^{-\alpha} \Bigg\rbrace\\
&-\bigg(\frac{1}{1+\beta}-\frac{1}{1-\alpha+\beta}\bigg)\Bigg],
\end{split}
\end{equation}

\noindent where $h_c$ is the disk scale-height at $r=r_{\rm c}$, disk angular velocity of $\Omega(r)\propto r^{-\alpha}$, and sound speed $c(r)\propto r^{-\beta}$. Equation \ref{eq:spiral} has been used to fit spiral arms observed in scattered light \citep{Muto2012, benisty2015}, although the approximations assumed to derive Eq. \ref{eq:spiral} may fail for massive planets \citep[$\gtrsim1~M_{\rm{Jup}}$, ][]{zhu2015}. This linear implementation results in 1 spiral for 1 planet, while in the non-linear case, one planet can generate one or more spirals \citep{zhu2015,Dong2015spirals}.

For Model 1, we fit Eq. \ref{eq:spiral} to the position of the maxima of F2. For this purpose we select the pixels of the S1 arm, masking out the inner ring. We also set $\alpha=1.5$ (Keplerian rotation), $\beta=0.45$ \citep[from the temperature profile in ][]{vanderMarel2015-isot}. The value of $h_c$ is not well constrained by any model of the system, and at the radii of interest it ranges between $0.08$ and $0.16$ \citep[e.g.][]{Andrews2011, Carmona2014, vanderMarel2015-isot}. Therefore, we fix the scale-height value to the average $h_{\rm c}=0.12$, so only two free parameters remain for the fit: $r_{\rm c}$ and $\theta_{\rm 0}$, which characterize the launching position of the spiral.

\begin{figure*}[!ht]
\epsscale{2}
\plotone{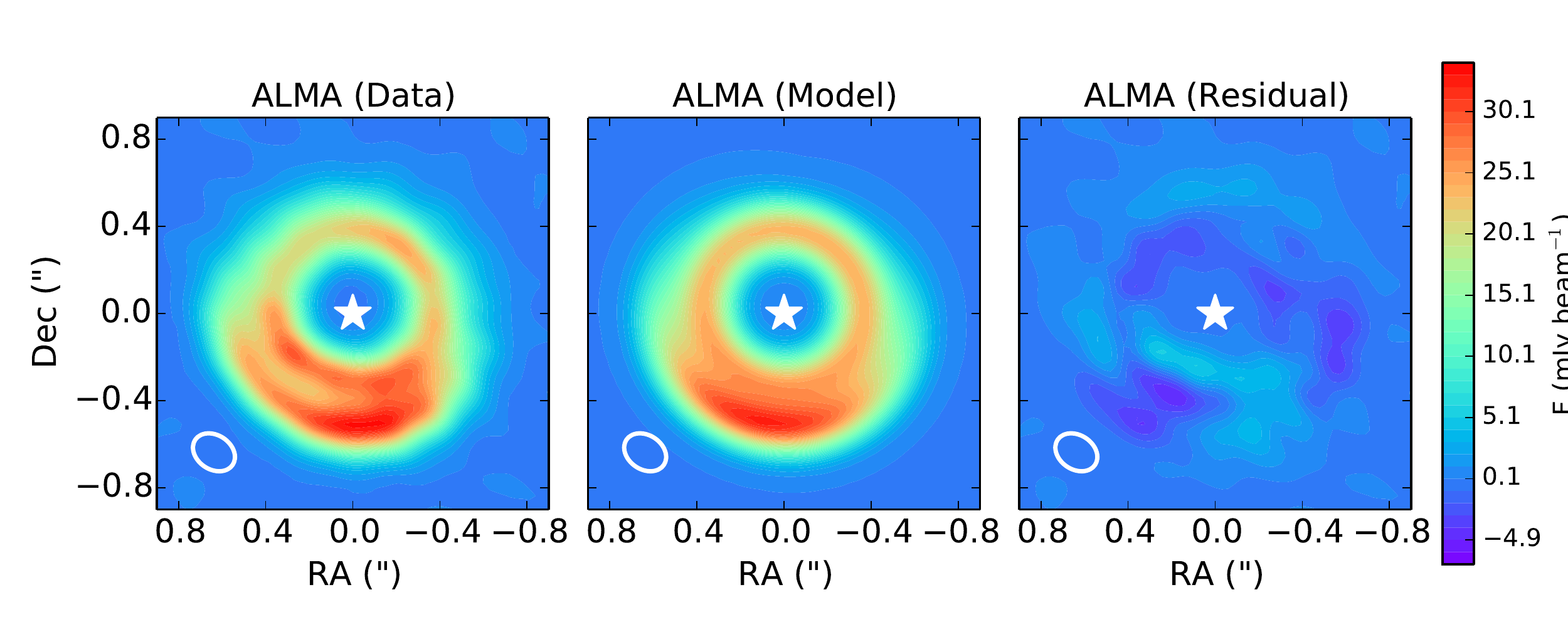}
\caption{The best-fit model for a ring in combination with a vortex (2D Gaussian) for the ALMA continuum data. Modeling has been performed in the uv plane.}
\label{fig:gaussfit}
\end{figure*}

We adopt an Orthogonal Distance Regression fitting procedure, that searches for the curve that minimizes the sum of the  distances to the data points orthogonally to the curve itself, thus assuming an observational error on both $\theta$ and $r$ in Equation \ref{eq:spiral}. We assume the uncertainty on the positions of the maxima to be equal to the FWHM of the beam. Finally, each data point is weighted proportional to the corresponding pixel intensity.

The fit in Figure \ref{fig:spiraloverlay} (left) shows that the F1 structure is mostly coradial {and hence the spiral launching position has to be very close to the central star ($r_{\rm c}< 0.2''$): the spiral pitch angle is close} to $0^{\circ}$. In such a scenario, F1 would be part of the 4th spiral winding. {However, the density waves after the first spiral winding are damped very efficiently due to the disk viscosity and pressure torque \citep[][]{Baruteau2014} and therefore this scenario is unrealistic to explain the observed azimuthal structure.} Figure \ref{fig:spiraloverlay} (right) shows that the ALMA continuum does not follow the best fit to the spiral arm in scattered light (blue dots).

\subsection{The Ring plus Asymmetry Model}
Model 2 describes the structure as a combination of a ring (F2) with a azimuthal asymmetry (F1). This model assumes that the asymmetry may originate from a vortex, using the vortex prescription by \citet{LyraLin2013} of a Gaussian in the radial and azimuthal direction:

\begin{equation}
F(r,\theta)=F_{\rm v}e^{-(r-r_{\rm v})^2/2\sigma_{\rm r,v}^2}e^{-(\theta-\theta_{\rm v})^2/2\sigma_{\rm \theta,v}^2},
\end{equation}

where $F_{\rm v}$ is the flux density at $(r_{\rm v},\theta_{\rm v})$, $r_{\rm v}$ and $\theta_{\rm v}$ (East of North) are the radial and azimuthal position of the asymmetry respectively, and $\sigma_{\rm r,v}$ and $\sigma_{\rm \theta,v}$ are the radial and azimuthal width of the asymmetry. F2, on the other hand, is modeled as a gaussian ring, 

\begin{equation}
F(r,\theta)=F_{\rm r}e^{-(r-r_{\rm r})^2/2\sigma_{\rm r,r}^2},
\end{equation}

\noindent where $F_{\rm r}$ is the flux density at $r_{\rm r}$, and where $r_{\rm r}$ and $\sigma_{\rm r,r}$ are the radial position and width of the ring respectively.

Our model has therefore 8 free parameters (5 for the asymmetry and 3 for the ring model), and we fit it to the {image} 
using the MCMC python package \texttt{emcee}. The chains from the fit show good convergence for all the free parameters, and the best fit parameters are: \\
\begin{tabular}{l|l}
\hline
$F_{\rm v}$&$1.44$ $\pm1.6\times10^{-3}$ mJy/pixel \\
$r_{\rm v}$&80.7 $\pm$  0.005 AU ($0.58'' \pm3.3''\times 10^{-5}$) \\
$\sigma_{\rm r,v}$&6.3 $\pm$ 0.008 AU ($0.045''\pm5.7''\times10^{-5}$) \\
$\theta_{\rm v}$&$172^{\circ} \pm0.02^{\circ}$ \\
$\sigma_{\rm \theta,v}$&$57^{\circ}\pm0.02^{\circ}$ \\
$F_{\rm r}$&$0.96$ $\pm6.1\times10^{-4}$ mJy/pixel \\
$r_{\rm r}$&51.3 $\pm$ 0.004 AU ($0.37'' \pm2.8''\times10^{-5}$) \\
$\sigma_{\rm r,r}$&8.1 $\pm$ 0.007 AU ($0.058''\pm4.8''\times10^{-5}$)\\
\hline
\end{tabular}

\begin{figure*}[!ht]
\epsscale{1}
\plotone{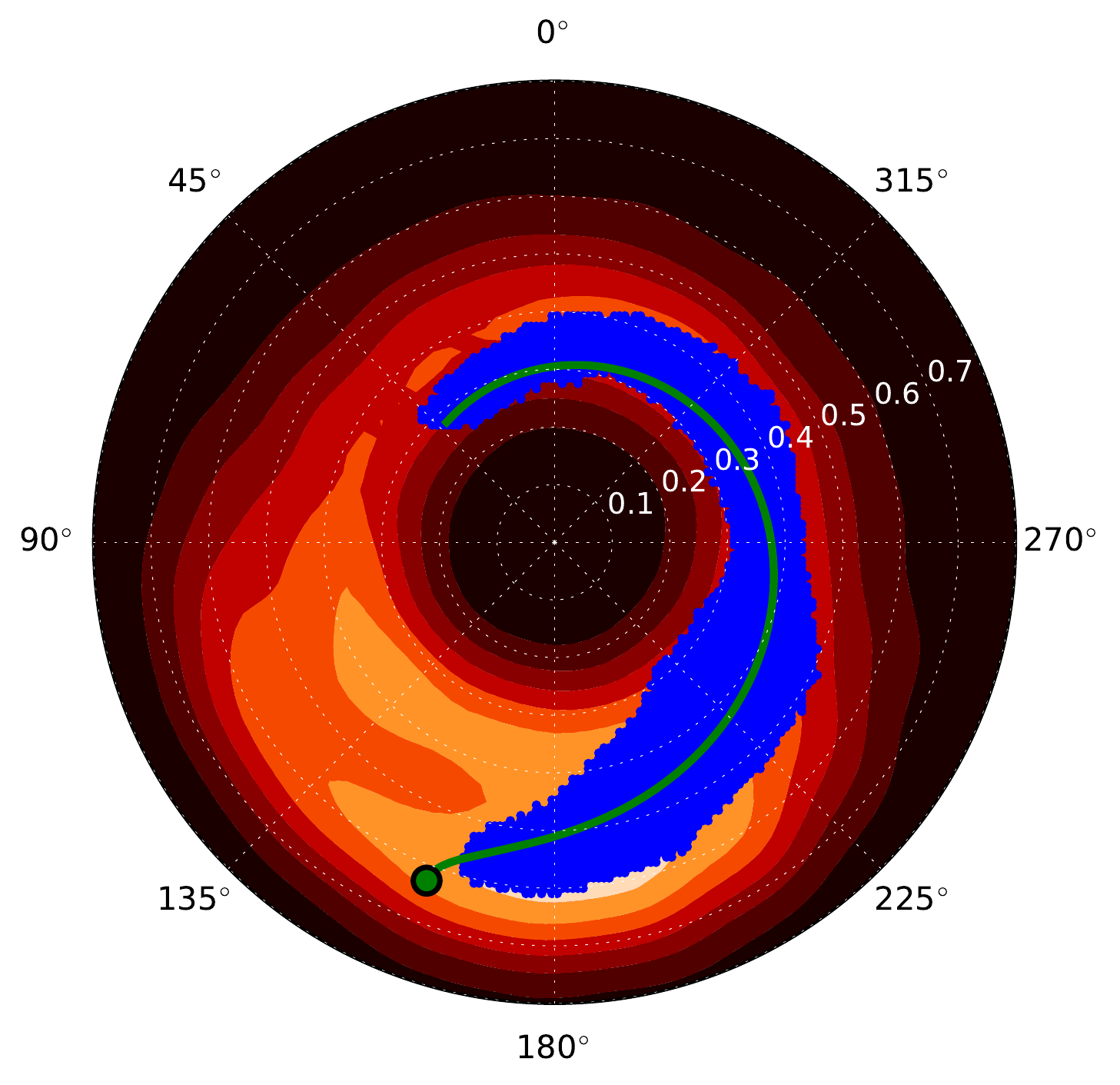}
\plotone{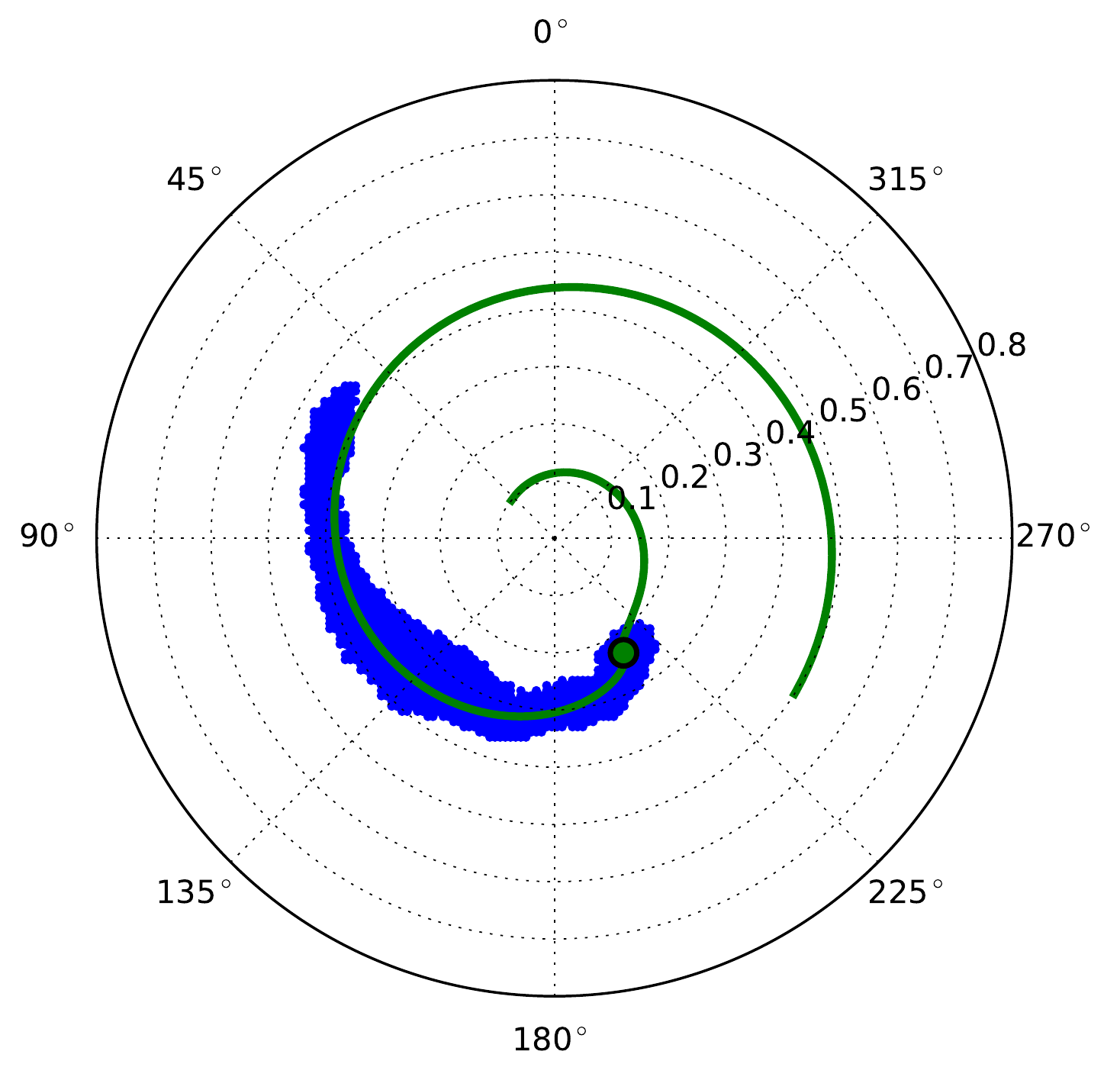}
\caption{{\bf Left.} Best fit for the S1 spiral in the scattered light data (deprojected), with a starting point inside the vortex. Overlaid on the ALMA image (colors), the blue dots indicate the data points of the PDI S1 feature with the central ring masked out, and the green line the best-fit spiral, with launching point $(r_c,\theta_0)=(0.62'',170^{\circ})$ marked as a circle. {\bf Right.} Best fit for the S2 spiral, with a starting point in the inner part of the disk. The blue dots are the brightest points of the PDI S2 feature and the green line the best-fit spiral, with launching point $(r_c,\theta_0)=(0.23'',211^{\circ})$ marked as a circle.}
\label{fig:fitS12}
\end{figure*}

The errors from the MCMC calculations are much smaller than the spatial uncertainty from the observations, which is typically $\sim10\%$ of the beam size (i.e. 2-3 AU). Figure \ref{fig:gaussfit} shows the comparison between the convolved model and the observations. The best fit was simulated onto the observed visibilities, and no significant differences were found with the convolved image. Some residuals are still present, mostly due to the asymmetry in the inner ring, but at the 10\% level of the original flux. The radius of the vortex is at a larger radius than found by earlier fitting of the millimeter data \citep{Perez2014,Pinilla2015beta}, which could be due to their central position being 11 AU away from this study.

\section{Discussion and conclusions}

\begin{figure*}[!ht]
\epsscale{2}
\plotone{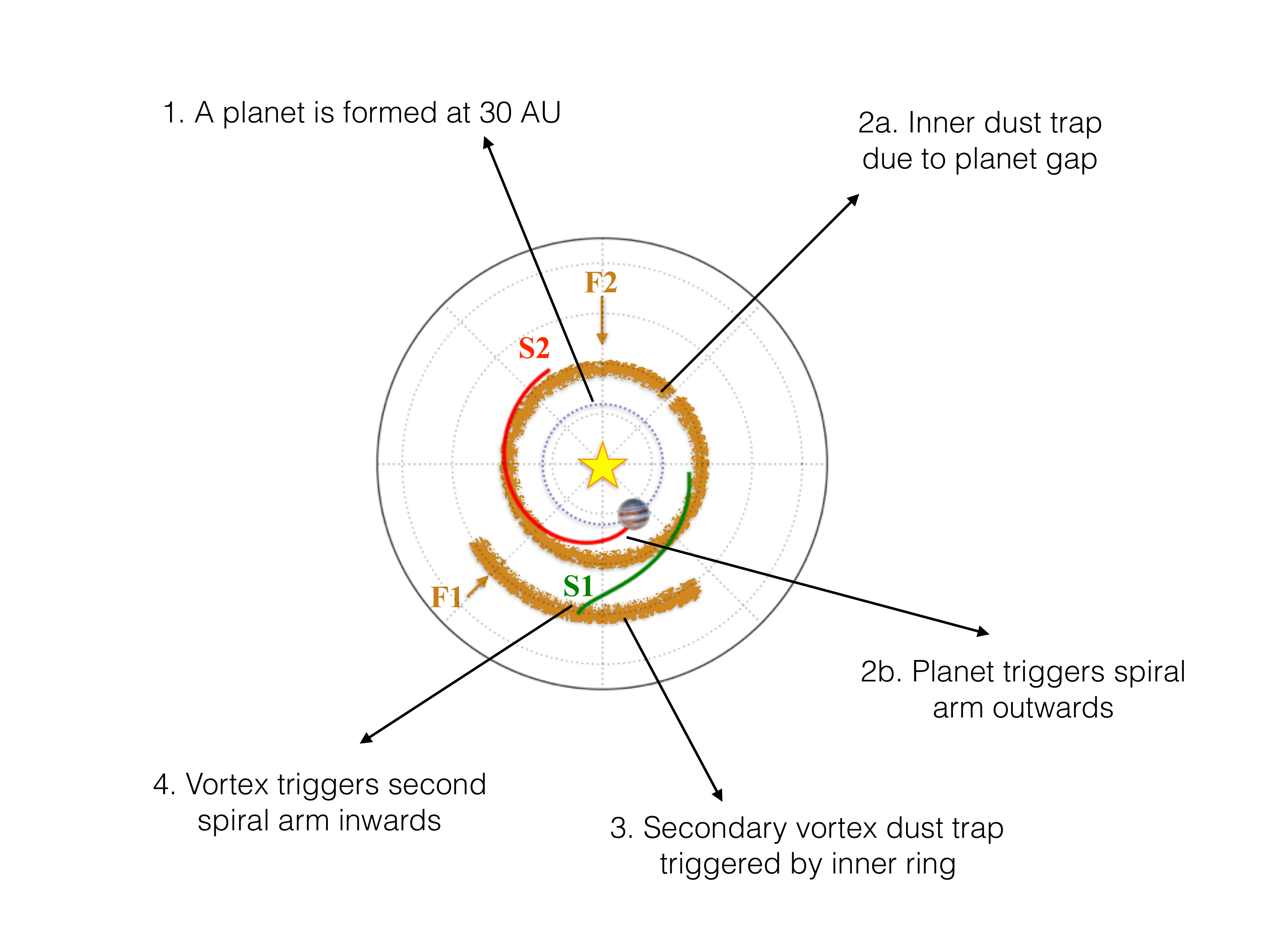}
\caption{Cartoon explaining the proposed scenario.}
\label{fig:cartoon}
\end{figure*}

The F1 feature is not consistent with the spiral arm prescription, but it can be described as a ring ($\sim50$ AU) with an asymmetry at $\sim80$ AU. Therefore we propose a new alternative scenario for this disk to explain the structure of both millimeter and scattered light data. The millimeter geometry is consistent with a model from \cite{Lobo-Gomes2015}, showing that a planet generates a pressure bump at $50$ AU (F2), which triggers a second generation vortex at $80$ AU (F1). The cavity radius of the gas and small grains \citep{Garufi2013,vanderMarel2015-isot} suggests the presence of a massive planet at 30 AU. A millimeter dust ring at 50 AU (F2) is consistent with this scenario, as the dust is trapped further out than the gas gap edge \citep{Pinilla2012b}. 

The ALMA and PDI data trace different grain size populations and disk heights, possibly driven by different mechanisms. However, it is striking that F1 coincides with the edge of the S1 arm. We propose that the S1 is triggered by a vortex that has created the dust asymmetry F1, since vortices can be massive enough to launch their own density waves in a disk when self-gravity is included in hydrodynamical models \citep[e.g.][]{BaruteauZhu2015}. Only a lower limit to the mass of the F1 feature can be set as the emission is partially optically thick: with a total flux of $\sim$200 mJy and a ISM gas-to-dust ratio of 100, the total mass is $>$16 M$_{\rm Jup}$ (using $M_{\rm gas} = 0.08*F_{\nu}(d/140 {\rm pc})^2$ M$_{\rm Jup}$, \citet{Cieza2008}). The outer extent of S1 (outside the vortex) remains undetectable in the PDI image due to the lower brightness in the outer disk.

\citet{Muto2012} find a best-fit for the launching point of S1 at $r_c$=0.39" (55 AU) at $\theta_0$=204$^{\circ}$, but with a large confidence interval (see Figure 5 in Muto et al.). Fitting the S1 spiral with an initial guess close to the center of the vortex results in the fit in Figure \ref{fig:fitS12}a with $r_c,\theta_0=0.6'',180^{\circ}$ (84 AU) and $h_c=0.08$. This launching point does not coincide exactly with the center of F1, although there is a large uncertainty due to the unknown scale height at this location. Furthermore, ALMA continuum observations trace the mm-dust, whose center may not coincide with the gas vortex \citep{BaruteauZhu2015}, and the vortex can be a large scale structure where the center of mass may not be well represented by a single location, contrary to a planet. 
 
On the other hand, the S2 spiral was best-fit by \citet{Muto2012} for $r_c,\theta_0$=0.9" (126 AU), 353$^{\circ}$, but we find that it can also be fit with a launching point in the inner part of the disk for $r_c,\theta_0$=0.23" (32 AU), 211$^{\circ}$ (Figure \ref{fig:fitS12}b). 
The launching point of S2 would be a massive planet, just inside the gas cavity radius \citep{vanderMarel2015-isot}. \citet{Stolker2016} finds a best fit for the S2 launching point to the VLT/SPHERE data slightly further in, at $r_c,\theta_0$=0.15" (21 AU), 247$^{\circ}$.

We propose that the combination of the scattered light and the millimeter observations is consistent with the following sequence of events (see Figure \ref{fig:cartoon}):

 \begin{enumerate}
\item A massive planet is formed at $\sim$30 AU radius. 
\item The planet triggers a spiral density wave outwards (PDI S2 feature).
\item The planet clears its orbit in the gas (CO observations) and creates a radial pressure bump at its edge where millimeter-dust gets trapped (ALMA continuum F2 feature).
\item The pressure bump creates an effective $\alpha$ viscosity that is large enough to induce accretion, depleting the gas and inducing a second pressure bump further out. The second pressure bump triggers RWI, forms a vortex and traps the millimeter-dust asymmetrically (ALMA continuum F1 feature).
\item The outer vortex triggers a spiral density wave inwards (PDI S1 feature).
\end{enumerate}

This scenario can potentially explain both PDI and millimeter observations.  Hydrodynamical models of gas and dust, including self-gravity, are required to check whether our proposed scenario can instead quantitatively explain the observed structures of HD135344B. 

One of the major uncertainties in the scenario are the fits to the locations of the launching points. The reason is that the scattered light data are mainly sensitive to changes in the scale height and therefore, the observed scattered light is significantly affected by geometric parameters. The observed spirals form only the illuminated inner part of a surface change. Also, the inner disk region may shadow the outer part and thus alter the intrinsic disk scale height distribution. In particular, the azimuthal angle of the continuum ALMA feature coincides with the brighter part of the closer-in S2 spiral and therefore, S2 may be casting a shadow on part of S1, affecting the fit of the launching points. 

Another caveat is the symmetry of the two spiral arms at the time of observation, suggesting a common nature such as proposed by \citet{Dong2015spirals} who demonstrates the trigger of two symmetric spiral arms by a single planet at 100 AU. As this planet has remained undetected, this scenario cannot be confirmed. On the other hand, if there are instead two launching points (32 and 86 AU), the two spirals would have distinct angular velocities and their symmetric appearance is fortuitous, making the scenario less probable. The orbital period of the 32 AU point is only 143 years, implying a 2.5$^{\circ}$/year angular shift. Repeating the scattered light observations in 5 years should clearly reveal the motion of this arm. If the asymmetry is indeed related to a vortex, an azimuthal shift of $\sim$0.1" (6$^{\circ}$) in the millimeter continuum (measurable at 0.2" resolution) is detectable after 10 years.

The scenario is an example of triggered planet formation, where the formation of a first planet can induce dust growth and potentially further planet formation in the outer disk.

\bibliographystyle{apj}



\acknowledgments
We are grateful to E.F.~van Dishoeck, M.~Tazzari, S.~Facchini and T.~Muto for useful discussions.  NM is supported by the Beatrice Watson Parrent Fellowship. Astrochemistry in Leiden is supported by NOVA, KNAW and EU A-ERC grant 291141 CHEMPLAN. This paper makes use of the following ALMA data: ADS/JAO.ALMA/2012.1.00158.S. ALMA is a partnership of ESO (representing its member states), NSF (USA) and NINS (Japan), together with NRC (Canada) and NSC and ASIAA (Taiwan), in cooperation with the Republic of Chile. The Joint ALMA Observatory is operated by ESO, AUI/NRAO and NAOJ. 




{\it Facilities:} \facility{ALMA}.

\end{document}